\documentclass[%
 aip,
amsmath,amssymb,
reprint,%
]{revtex4-1}

\usepackage{graphicx}
\usepackage{dcolumn}
\usepackage{bm}
\usepackage{hyperref}
\hypersetup{
    colorlinks = true,
    urlcolor   = blue,
    citecolor  = blue,
    linkcolor = blue,
}
\usepackage[utf8]{inputenc}
\usepackage{csquotes}
\usepackage[T1]{fontenc}
\usepackage{mathptmx}
\usepackage{etoolbox}
\usepackage{color}
\usepackage[mathlines]{lineno}
\usepackage[table, dvipsnames]{xcolor}
\usepackage{rotating}
\usepackage[export]{adjustbox}
\usepackage{array}
\usepackage{devanagari}

\makeatletter
\def\@email#1#2{%
 \endgroup
 \patchcmd{\titleblock@produce}
  {\frontmatter@RRAPformat}
  {\frontmatter@RRAPformat{\produce@RRAP{*#1\href{mailto:#2}{#2}}}\frontmatter@RRAPformat}
  {}{}
}%
\makeatother
\begin{document}
\preprint{AIP/123-QED
}

\title[]{Insights of Transitions to Thermoacoustic Instability in Inverse Diffusion Flame using Multifractal Detrended Fluctuation Analysis}
\author{Somnath De}
\email{somnathde.mec@gmail.com}
\affiliation{Department of Mechanical Engineering, Institute of Engineering and Management, Kolkata, India 700091}%
\author{Soham Bhattacharya}%
\affiliation{Department of Mechanical Engineering, Politecnico di Milano, Via Raffaele Lambruschini, 20156 Milano MI, Italy}%

\author{Arijit Bhattacharya}%
\affiliation{Department of Aerospace Engineering, Indian Institute of Technology Kanpur, Kanpur, Uttar Pradesh, India 208016}%

\author{Sirshendu Mondal}%
\affiliation{Department of Mechanical Engineering, NIT Durgapur, India  713209}%

\author{Achintya Mukhopadhyay}%
 \email{achintya.mukhopadhyay@jadavpuruniversity.in}
\affiliation{Department of Mechanical Engineering, Jadavpur University, Kolkata, India 700032}%

\author{Swarnendu Sen}%
\affiliation{Department of Mechanical Engineering, Jadavpur University, Kolkata, India 700032}%
\date{\today}

\begin{abstract}
The inverse diffusion flame (IDF) can experience thermoacoustic instability due to variations in power input or flow conditions. However, the dynamical transitions in IDF that lead to this instability when altering control parameters have not been thoroughly investigated. In this study, we explore the control parameters through two different approaches and employ multifractal detrended fluctuation analysis to characterize the transitions observed prior to the onset of thermoacoustic instability in the inverse diffusion flame. Our findings reveal a loss of multifractality near the region associated with thermoacoustic instability, which suggests a more ordered behavior. We determine that the singularity exponent, the width of the multifractal spectrum, and the Hurst exponent are reliable indicators of thermoacoustic instability and serve as effective classifiers of dynamical states in inverse diffusion flames.

\end{abstract}


\maketitle

\section{Introduction}\label{sec:Introduction}
\textquotedblleft Rijke tube\textquotedblright~\cite{rijke1859lxxi,rijke1859notiz} discovered by Professor P.L. Rijke in 1859, is a cylindrical tube with open ends that contains a heat source. This unique configuration generates two distinct types of flames, influenced by the introduction of oxidizer and fuel: normalized diffusion flame (NDF)~\cite{hunger2017comparative} and inverse diffusion flame (IDF)~\cite{de2017numerical,mikofski2006flame}. In the NDF configuration, fuel is injected at the center, surrounded by an oxidizer. In contrast, the IDF configuration has a coaxial setup where the fuel surrounds a centralized flow of oxidizer. While both configurations provide stable operations, IDF is generally considered more suitable for practical applications because it results in lower emissions of thermal $NO_x$ and soot~\cite{sze2006appearance, choy2012pollutant, kotb2016comparison,kotb2016comparison}, is less susceptible to lean blowout~\cite{sobiesiak2005characteristics}, and offers a wider stability margin~\cite{zhen2010thermal}. In IDF configuration, the central air jet stabilizes the flame and enhances flame stability by creating negative pressure along the flame axis.

An IDF consists of two distinct zones: the flame torch and the flame base. These zones are connected by a section called the flame neck. In the flame neck region, intense mixing occurs due to the controlled velocity of the air jet, allowing air and unburned fuel to mix effectively before entering the flame torch zone~\cite{miao2016effect}. Furthermore, an IDF combines the benefits of both normal diffusion flames and premixed flames. Therefore, the IDF concept can offer an alternative approach to both lean premixed combustion and normal diffusion flames in applications such as rocket engine combustion, staged combustion, industrial burners, and aviation~\cite{zhen2021state}. IDF is also beneficial in heating applications where a single flame needs to have a higher thermal load (i.e., a higher heat transfer rate)~\cite{zhen2021state}. Compared to typical diffusion flames or premixed flames, IDF is cleaner and safer for these types of applications.
 
However, the IDF configuration converts heat into sound by promoting self-sustaining and self-amplifying standing waves.
This phenomenon is highly relevant to the safety of combustion chambers in industrial burners~\cite{sujith2020complex}, rocket propulsion ~\cite{de2024feedback, hashimoto2019spatiotemporal, kasthuri2022coupled}, etc. The presence of high-amplitude acoustic waves from thermoacoustic instability can damage the structural integrity of engines~\cite{sujith2020complex}. This instability often manifests as loud noise, known as combustion roar, which is undesirable in both industrial and consumer applications and can lead to mission failures. ~\cite{delaat2008characterization}. Additionally, thermoacoustic isntability can disrupt energy conversion efficiency, resulting in higher fuel consumption and lower output power. In some cases, it may also increase pollutant emissions, particularly in industrial combustion systems. Thus, the challenges of thermoacoustic instability in gas turbine engines, rocket motors, industrial boilers, oxygen-fuel Combustion and industrial process control remain significant for researchers in the twenty-first century~\cite{sujith2021thermoacoustic,silva2023intrinsic,morgans2024thermoacoustic}. Consequently, investigating thermoacoustic instability across various engine combustors, power generation systems, and aerospace applications has become a key area of focus for scientists. 
In this study, we examine the critical issues of thermoacoustic instability in an IDF combustor. 

Earlier, a significant amount of research had been conducted on IDF to understand their characteristics. Recently, Sun et al.~\cite{sun2024numerical} employed the Reynolds-averaged Navier-Stokes (RANS) numerical method to analyze the combustion performance of an IDF combined with air swirling. Their results demonstrated that the swirling action creates a central recirculation zone (CRZ) and two external recirculation zones (ERZ) at the head of burner, which contributes to stabilizing combustion. 
Due to the wide variety of physical and chemical properties associated with IDF, previous research was often multi-faceted, focusing on key aspects, including burner design, flow-flame operation, flame appearance (such as size, color, and shape), and flame structure~\cite{hunger2017comparative, miao2014effect, patel2019effect, zhu2018near}. Further, several studies~\cite{mahesh2008flame, mahesh2010flame} have examined the stability and emission characteristics of dual flame structures, focusing on centerline temperature distribution and oxygen concentration in turbulent IDF. The combustion characteristics of IDF are heavily influenced by fuel/air mixing, making the fluid dynamics of the flame a crucial factor. In the laminar regime of IDF, fuel and oxidizers are transported to the reaction zone primarily through diffusion. In contrast, in turbulent regimes, the mixing of fuel and air occurs more vigorously due to fuel impingement and radial and tangential flows. As a result, IDF exhibits several dynamic characteristics. While many studies have been conducted on IDF combustors, their dynamics remain unexplored.     

Nalawade et al.~\cite{nalawade2023dynamics} attempted to investigate the blowoff dynamics of turbulent IDF by synchronizing high speed CH* chemiluminescence imaging with pressure sensors. They found that blowoff happens due to higher strain rates caused by the simultaneous spatiotemporal interaction of turbulent flow structure with flame surface. Kabiraj et al. ~\cite{kabiraj2012nonlinear} examined the nonlinear transitions from thermoacoustic oscillations to flame blowout due to intermittency, resulting from changes in the position of the combustion source within a ducted laminar premixed conical flame. They observed a subcritical Hopf bifurcation in the mean acoustic pressure oscillations as the combustor transitions into limit cycle oscillations. Additionally, the limit cycle oscillations displayed a bifurcation into quasi-periodic oscillations, leading to an intermittency state. Thus, the study ~\cite{kabiraj2012nonlinear} explored the pathway to blowout via thermoacoustic instability in a single conical flame burner confined within a closed–open borosilicate glass duct.

Further, Sen et al.~\cite{sen2018dynamic} investigated the dynamics of a laboratory-scale ducted inverse diffusion flame with the variation of the position of flame within the duct. Their study revealed the presence of limit cycles, type-II intermittency, and homoclinic orbits through phase space reconstruction and recurrence analysis on the time series of acoustic signals. Recently, Bhattacharya et al.~\cite{bhattacharya2021recurrence} investigated the onset of thermoacoustic instability using three recurrence network tools: global efficiency, average degree centrality, and global clustering coefficient.
However, 
unlike the extensive research on premixed and non-premixed flames, transitions to thermoacoustic instability in IDF have received little attention which underscores the need for further research into the dynamics of IDF~\cite{zhen2021state} with varying air and fuel flow rates.

The dynamics associated with thermoacoustic instability are inherently non-linear, making it essential to investigate these transitions using non-linear methods~\cite{sujith2020complex,sujith2021thermoacoustic}. Illingworth et al.~\cite{illingworth2013finding} tried to investigate the nonlinear thermoacoustic oscillations of a ducted Burke-Schumann diffusion flame by employing a combination of matrix-free continuation and flame-describing function (FDF) methods. Their study revealed that the inlet boundary conditions during the thermoacoustic cycle significantly contribute to the nonlinearity observed in the flame's response to harmonic velocity fluctuations across a range of forcing frequencies and amplitudes. However, it is important to note that the flame describing function serves as an approximation, linearizing the nonlinear interactions between acoustic waves and flame dynamics. This simplification may neglect critical elements of complex flame behavior. On the other hand, the complexity of unsteady combustion behavior arises from the presence of multiple spatial and temporal scales. As a result, the fractal dimension of such systems can provide valuable insights into their complexity~\cite{raghunathan2020multifractal}. 
Typically, a single dimension can effectively describe the fractal dynamics of a system, leading to the concept of "monofractal." The idea of self-similarity, which is used to represent similar dynamics across different temporal and spatial scales in monofractals, was first introduced by Mandelbrot~\cite{mandelbrot1999intermittent} in the study of turbulent flows. However, in complex systems, understanding their multiscale behavior often requires a range of fractal dimensions. This approach examines the spectrum of fractal dimensions and is known as multifractal analysis~\cite{kantelhardt2002multifractal}.
Mandelbrot first introduced the concept of multifractality in 1983~\cite{mandelbrot1983fractal} and further developed it in 1989~\cite{mandelbrot1989multifractal}. Since then, the popularity of this concept has spread rapidly in various fields, including engineering~\cite{meneveau1991multifractal}, finance~\cite{morales2014dependency}, geophysics~\cite{telesca2006measuring}, astrophysics~\cite{jones1988multifractal}, chemistry~\cite{saha2020multifractal}, and medical science~\cite{huynh2023multifractality,he2022multifractal}. 


\begin{figure*}[ht!]
\includegraphics[width=1.55\columnwidth]{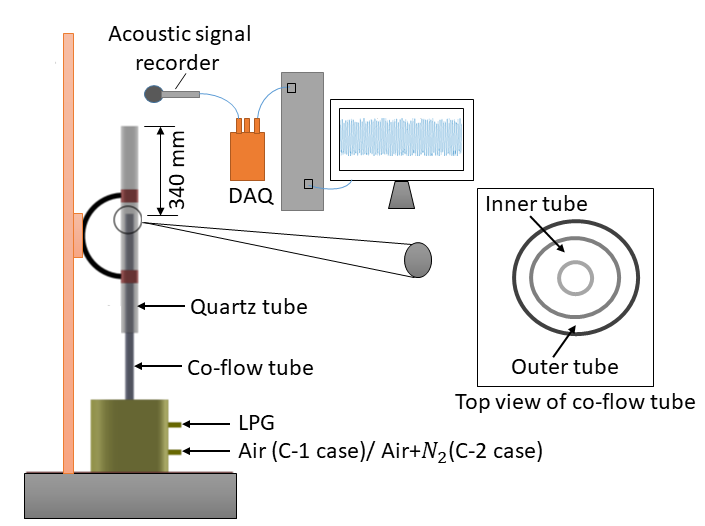}
\caption{Here, we introduce the experimental setup of Rijke tube in which the experiments of inverse diffusion flame (IDF) are conducted. The setup features a co-flow tube, with air supplied through the inner tube and fuel through the outer tube. A quartz tube serves as the wall of the combustion zone. Additionally, an acoustic signal from the combustor is recorded using a data acquisition system.
}\label{fig:setup}
\end{figure*}

Like different fields, in the combustion field, the multifractal approach has received significant attention. Niu et al.~\cite{niu2008multifractal} confirmed the multifractal nature of the flame in a four-burner impinging entrained-flow gasifier by performing multifractal detrended fluctuation analysis on the high-speed flame images. Later, Sen et al.~\cite{sen2008multifractal} investigated the dynamics of the cycle-to-cycle heat release rate variations in spark ignition engines at different spark advance angles and found a higher degree of multifractality of the dynamics at higher spark advance angles. In their subsequent study, Sen et al.~\cite{sen2010analysis} characterized the changes in the dynamics of a lean-fueled, multi-cylinder spark-ignition (SI) engine using multifractals and wavelets. Further, Curto-Risso et al.~\cite{curto2010monofractal}
used both monofractal and multifractal approaches to characterize the heat release rate fluctuations for different fuel-air ratios in a simulated spark ignition engine. They identified the presence of weak anticorrelations in the fluctuations for lean and stoichiometric conditions. Gotoda et al.~\cite{gotoda2012characterization} found the multifractal behavior of acoustic pressure oscillations in a swirl-stabilized turbulent combustor prior to the occurrence of a lean blowout. Nair and Sujith~\cite{nair2014multifractality} noticed a scale invariance and multifractal behavior of unsteady pressure oscillations during the presence of combustion noise state in both swirl and bluff body stabilized combustors. They found that the multifractal behavior disappears at the onset of thermoacoustic instability in those combustors. Recently, Ragunathan et al.~\cite{raghunathan2020multifractal} studied the spatiotemporal dynamics of a bluff body stabilized combustor by performing multifractal analysis and discovered that multifractality exists in the flame topology for all dynamical phases during the transition to thermoacoustic instability. They found the fluctuations in  multifractality spectrum at the onset of thermoacoustic instability that is nearly similar to the observation noticed by Sujith and Nair~\cite{nair2014multifractality} in the temporal dynamics.

Similar to these works mentioned in the last paragraph on conventional turbulent premixed combustion~\cite{gotoda2012characterization, nair2014multifractality,raghunathan2020multifractal}, the change in flow conditions or power input can
induce thermoacoustic instability in IDF~\cite{sen2018dynamic}. However, in contrast to conventional flames, the transitions leading to thermoacoustic instability in IDF caused by changes in control parameters are not well understood. Additionally, unlike the laminar and turbulent premixed flames~\cite{kabiraj2012nonlinear}, the development and testing of strategies for predicting the onset of thermoacoustic instability in IDF is limited.

Hence, in light of the advancement in fractal theory, we apply multifractal detrended fluctuation analysis to examine the transitions to thermoacoustic instability in an IDF system under two different conditions: A. a constant power input (i.e., fuel flow rate) while Reynolds number is varied from laminar to turbulent. B. constant flow condition while power input is varied. We identify different regimes based on the variation in acoustic emissions captured by a microphone. Our analysis reveals a higher degree of multifractal behavior in the system dynamics at lower Reynolds numbers (with constant power input) and at lower power inputs (with constant Reynolds numbers).  A loss of multifractality is captured when the system transitions to a higher Reynolds number (constant power input) and higher power input (constant Reynolds number). Further, we find that the Hurst exponent obtained through monofractal and multifractal spectrum tools, such as spectrum width and singular strength, are more effective than the conventional recurrence network matrices~\cite{bhattacharya2021recurrence} in characterizing the dynamical regimes of the IDF system near thermoacoustic instability.


The organization of the paper is as follows. In sec.~\ref{sec:Experimental facilities}, we present the details of experimental facilities and the acquisition of data. We explain the steps used in the multifractal detrended fluctuation analysis (MFDFA) method for the present analysis in sec.~\ref{sec:method}. We discuss the findings by performing a multifractal approach on the data of IDF in sec.~\ref{sec:result}. Finally, we summarize the significant outcomes from this study in Sec.~\ref{sec:conclusions}.


\section{Experimental facilities}\label{sec:Experimental facilities}

\begin{table*} 
 \caption{The operating conditions for two case studies, C-1 and C-2, are detailed in the table below. In case study C-1, the Reynolds number ($Re$) varies from laminar to turbulent flow, ranging from 1566 to 5950, while the power input remains constant at 0.39 kW. In case study C-2, the power input varies between 0.50 kW and 0.68 kW, while the Reynolds number is kept nearly constant.}\label{Tab:1}
\centering
\setlength\arrayrulewidth{1.0 pt}
\begin{tabular}{p{3.4cm}|p{3.4cm}|p{3.4cm}|p{3.4cm}}
\hline
\begin{flushleft} Case studies \end{flushleft}&\begin{flushleft}$Re$\end{flushleft} & \begin{flushleft}Power Input (PI) in KW \end{flushleft} & \begin{flushleft} Global equivalence ratio ($\Phi$)\end{flushleft}\\
\hline
C-1 & $\sim$ 1566 - 5950  & 0.39 & 0.37-0.098\\
C-2& $\sim$ 5116–5168 & 0.50-0.68 & 0.2  \\
\hline
\end{tabular}
\end{table*}

The experimental setup consists of a vertically positioned open-open Rijke tube, as shown in Fig.~\ref{fig:setup}. This tube is made of quartz, with a total length of 800 mm, an inner diameter of 60 mm, and a thickness of 2.5 mm. The position of the flame within the tube can be adjusted; however, for this study, the flame location (the burner tip) is fixed at 340 mm from the top end of the quartz tube. The co-flow burner comprises two concentric tubes. The inner tube has a diameter of 14 mm and an outer diameter of 16 mm. The outer tube has an inner diameter of 28.6 mm and an outer diameter of 34 mm. Air is supplied through the inner tube, while LPG (Liquid Petroleum Gas, which consists of 60$\%$ propane and 40$\%$ butane) is provided through the annular space between the inner and outer tubes.

The fuel and air are supplied at ambient pressure and room temperature (298 K). We use a mass flow controller (make: Aalborg, range: 0–10 lpm $CH_4$, calibrated for LPG, accuracy: 1$\%$ of full scale reading) to regulate the fuel flow rate ($\dot{Q}_{F}$), while an air mass flow controller (make: Alicat, range: 0–250 lpm, accuracy: 0.8$\%$ of reading + 0.2$\%$ of full-scale reading) is used to regulate the air flow rate ($\dot{Q}_{A}$). We introduce $N_2$ in case C-2 (see Table~\ref{Tab:1}) to compensate for changes in the reactant mixture flow rate when maintaining approximately constant Reynolds number ($Re = \frac{4\dot{Q}_{A} \rho_{A}}{\pi \mu_{A} d_{i}}$, where, $\rho_{A}$ = density of air, $\mu_{A}$ = dynamic viscosity of air, $ d_{i}$ = diameter of inner tube) and equivalence ratio ($\Phi = \frac{(\dot{Q}_{F}/\dot{Q}_{A})_{act}}{(\dot{Q}_{F}/\dot{Q}_{A})_{st}}$, $act$ and $st$ refer to actual and stoichiometric conditions respectively). $N_2$ is supplied from a pressurized cylinder and metered using a mass flow controller (make: Aalborg, range: 0–60 lpm $CH_4$, calibrated for $N_2$, accuracy: 1$\%$ of full-scale reading). $N_2$ and air are mixed in a header before being injected through the inner tube as an inner oxidizer jet. In case C-1 (Table~\ref{Tab:1}), we change $Re$ and $\Phi$ while maintaining a constant power input  ($PI$).
We position a Philips SBCMD110/01 audio recorder near the top end of the Rijke tube. The audio recorder is connected to a data acquisition (DAQ) system, enabling us to collect the acoustic signals generated by the Rijke tube at a sampling frequency of 44.1 kHz.

\section{Methodology}\label{sec:method}

The multifractal detrended fluctuation analysis (MFDFA)~\cite{kantelhardt2002multifractal} can be implemented using the following steps:
Let us consider a stochastic event, represented by a time series, x, with a length of N.
\\ \textbf{Step 1: \textit{Subtraction of mean}:}

MFDFA focuses on analyzing the characteristics of fluctuations in stochastic events. By subtracting the mean over whole time series~\cite{li2019novel}, we create a modified time series ($x^{\prime}$) with a zero mean. This approach provides a clearer understanding of the profile of the time series. 

The new modified series, $x^{\prime}$, is defined as:
\begin{equation}
x^{\prime}(i) = x_i - \frac{\sum_{j=1}^{N} x_j}{N},
\label{eq:1}
\end{equation}
\\\textbf{Step 2: \textit{Determining nature of time series}:}

In general, MFDFA is employed to treat time series that are random walk in nature. This technique uncovers essential patterns, making it crucial for informed decision-making in data analysis. If the time series shows a noise-like structure instead of resembling a random walk, it must be transformed into a series that displays random walk characteristics before applying MFDFA.
Integrating a noise-like series yields a random walk-like series that can then be treated with MFDFA for further analysis~\cite{ihlen2012introduction}.
If $x^{\prime}$ is noise-like in nature, then, supposing $Z$ is the random walk derived from $x^{\prime}$: 
\begin{equation}
Z_i = \sum_{k=1}^{i} x^{\prime}_k,
\label{eq:2}
\end{equation}
It should be noted that the first step of this process, i.e., the mean subtraction, is optional for a random walk-like series. However, it is compulsory to implement the same for noise-like series before performing the integration to convert it into a random walk. 
To distinguish between random-walk and noise-like time series, a potential solution proposed by Eke et al.~\cite{eke2002fractal} is to conduct a monofractal DFA prior to performing MFDFA. Time series exhibit noise-like characteristics when the Hurst exponent, $H$ (discussed later in the current section), falls between 0.2 and 0.8. In this scenario, MFDFA can be applied directly without the need for transformation of the time series. Conversely, when $H$ is within the range of 1.2 to 1.8, the time series behaves more like a random walk. 
\\\textbf{Step 3: \textit{Determining scales}:} 

At this step, MFDFA constructs a statistical measure known as the structure function by dividing the random walk time series, $Z$, into non-overlapping segments based on a range of scales determined by the researchers according to phenomenological considerations and variables. 
In addition to the range of scales, it is important to specify the number of scales for which the structure function is to be constructed. Generally, the greater the number of scales, the more reliable the calculation becomes. However, this also leads to an increase in computational cost. 
Let the selected maximum scale, minimum scale, and number of scales be $S_{max}$, $S_{min}$ and $m$, respectively. Then the scales are given by $\textbf{S}$ = $\{S_{min}$=$S_1$, $S_2$, $S_3$, … , $S_{m-1}$, $S_m$ = $S_{max}\}$. 
It is sensible to use equal logarithmic spacing between the scales instead of linear spacing as this ensures optimal fitting while determining the Hurst exponent (which is discussed later) from the slope of structure-function versus scale on a log-log plot. 
Then,
\begin{equation}
S_i = S_{min} + i(\frac{logS_{max} - logS_{min}}{m-1}),
\label{eq:3}
\end{equation}
where, $i$ = 2, 3, …, $m-1$. 
$S_{min}$ and $S_{max}$ are taken as $2^{9}$ and $2^{14}$.  
According to the thumb rule, $S_{max}$
is kept less $N$/10 to ensure at least 10 segments of $Z$ are formed, and MFDFA works accurately ~\cite{ihlen2012introduction}.
\\\textbf{Step 4: \textit{Calculation of structure-function}:}


For each of the scales, $S_{i}$, the random walk, $Z$ or the integrated time series is divided into $k_{i}$ number of non-overlapping segments or windows of the length $S_{i}$, such that:

\begin{equation}
k_i \equiv \lfloor \frac{N}{S_i} \rfloor
\label{eq:4}
\end{equation}
The floor function is used to obtain the greatest integer that is less than or equal to $\frac{N}{S_i}$.

Let $N_k$ denote the elements of the $k^{th}$ segment, then, the resulting segments corresponding to scale $S_{i}$ are:
$(N_1)_i$= $Z_1$, $Z_2$, $Z_3$,…, $Z_{S_i}$\\
$(N_2)_i$ = $Z_{S_{i}+1}$, $Z_{S_{i}+2}$, $Z_{S_{i}+3}$,…, $Z_{2S_i}$\\
$(N_3)_i$  = $Z_{2S_i+1}$, $Z_{2S_i+2}$, $Z_{2S_i+3}$,…, $Z_{3S_i}$\\
…\\
$(N_k)_i$ = $Z_{(k-1)S_i +1}$, $Z_{(k-1)S_i+2}$, $Z_{(k-1)S_i + 3}$,…, $Z_{kS_i}$\\
where $i$ = 1, 2, …, $m$

The length of the time series, $N$, in general, is not an integral multiple of the scales, $S_i$. As a result, a small segment of the series, $Z_c$, $Z_{c+1}$, …, $Z_n$, where $c = (S_i \times k_i)+1$ and $m – k_i < S_i$ remains unexplored. In order to avoid ignoring this small segment, the same process is repeated starting from the end of the series or by reversing the series. As a result, for each scale, $S_i$, $2k_i$ number of segments are formed
from the reversed series given by:
\\$(N_{k+1})_i$= $Z_N$, $Z_{N-1}$, $Z_{N-2}$,…, $Z_{N-S_i}$\\
$(N_{k+2})_i$ = $Z_{(N-S_i)-1}$, $Z_{(N-S_i)-2}$, $Z_{(N-S_i)-3}$, …, $Z_{(N-2S_i)}$\\
$(N_{k+3})_i$ = $Z_{(N-2S_i)-1}$,$Z_{(N-2S_i)-2}$, $Z_{(N-2S_i)-3}$, …, $Z_{(N-3S_i)}$\\
…\\
…\\
$(N_{2k})_i$ = $Z_{(k-1)S_i +1}$, $Z_{(k-1)S_i+2}$, $Z_{(k-1)S_i + 3}$,…, $Z_{kS_i}$
where $i$ = 1, 2, …, $m$

For each of the segments, 
the local trend is calculated via least square method to generate the fitting polynomial in the $k^{th}$ window, $y_{k}$. This polynomial can be linear, quadratic or of higher orders and accordingly the MFDFA removes trends of degree one, two or higher and the technique is then labelled as MFDFA1, MFDFA2 and so on. 
The polynomial, $y_{k}$, is then subtracted from elements of the random walk, $Z$, belonging to the $k^{th}$ segment to complete the de-trending. The variance, $F^2$, is then calculated for $k^{th}$ segment and scale, $S_i$ as:
\begin{equation}
F^2 (k,S_i)=  \frac{1}{S_i}\sum_{j=1}^{S_i} \{Z[(k-1)S_i+j]-y_k(j)\}^2, 
\label{eq:5}
\end{equation}

Next, the $q$-orders are used to provide weightage for the local fluctuations in order to identify the self-similarity of the fluctuations at different time scales. The $q$-orders must have both positive and negative values that are not necessarily integers. The negative $q$-orders provide higher weightage to small-scale variations, while the positive $q$-orders render structure functions with higher weightage to large-scale variations. However, using extreme $q$-orders can lead to the dominance of fluctuations, which results in skewed and imprecise calculations. Therefore, it is important to be cautious when selecting the $q$-orders for a study. In most combustion studies, a $q$-order range of -5 to 5 has been commonly used~\cite{kasthuri2019dynamical}.

The qth order fluctuation function or the structure-function, $F_q$, is calculated for each $q$-order, by averaging the variance, $F^2$, for all $2k_i$ segments constructed from each scale $S_i$ as:
\begin{equation}
F_q (q,S_i)=\{\frac{1}{2k_i}\sum_{k=1}^{2k_i}[F^2 (k,S_i)]^{\frac{q}{2}}\}^{\frac{1}{q}}.
 \label{eq:6}
\end{equation}
\\\textbf{Step 5: \textit{Introduction of Hurst exponent:}}

For each $q$-order, $F_q$ is related to the scales $S_i$, according to power-law if the series is long-range correlated and has elements of self-similarity:
\begin{equation}
F_q (q,S_i)\sim S_i^{H_q}
\label{eq:7}
\end{equation}
On a log-log plot, 
scales $S_i$ (where $i$ = 1, 2, …, $m$)  are plotted on the abscissa and the structure function corresponding to each of these scales is plotted on the ordinate for the various $q$-orders considered. A linear best fit line is generated via least square method and the slope of this line gives us the generalized Hurst exponent($H(q)$) for various $q$. If $H(q)$ is dependent on $q$, the signal has multifractal characteristics. For positive values of $q$, $H(q)$ represents the scaling behavior of segments with large fluctuations, while it represents the scaling behavior of segments with small fluctuations for negative values of $q$. For a specific case where $q=2$, the term $H_q$ is commonly called the Hurst exponent~\cite{hurst1951long}.
\\\textbf{Step 6: \textit{Calculation of multifractal parameters}:} 

Using the generalized Hurst exponent, $H(q)$, the other MFDFA parameters are constructed.
The scaling exponent or mass exponent, $\tau(q)$ is calculated as: $\tau(q) = q H(q)-1$.
The singularity spectrum, $f(\alpha)$ is constructed from the scaling exponent as they are related via the Lengendre transform~\cite{zia2009making}:
\begin{equation}
\alpha=\frac{\partial\tau(q)}{\partial q}
\label{eq:8}
\end{equation}
\begin{equation}
f(\alpha) = q\alpha-\tau(q)     
\label{eq:9}
\end{equation}
Here, $\alpha$ is called the singularity strength. Higher values of $\alpha$ signifiy the smaller degree of singularity. On the contrast, smaller vales of $\alpha$ imply larger degree of singularity. $\alpha = 1$ indicates the uniform distribution of time series. Next, a multifractal spectrum can be obtained by plotting $f(\alpha)$ against $\alpha$ for a comprehensive representation. Many natural and artificial systems, such as financial markets, geological formations, climate patterns, and turbulence are highly irregular and exhibit non-uniform scaling properties. The multifractal spectrum reveals how these systems 
are structured and exhibit different levels of complexity at different spatial or temporal scales. 
This is crucial for capturing the nuances that a single fractal dimension cannot provide.  

\begin{figure*}[ht!]
\includegraphics[width=1.8\columnwidth]{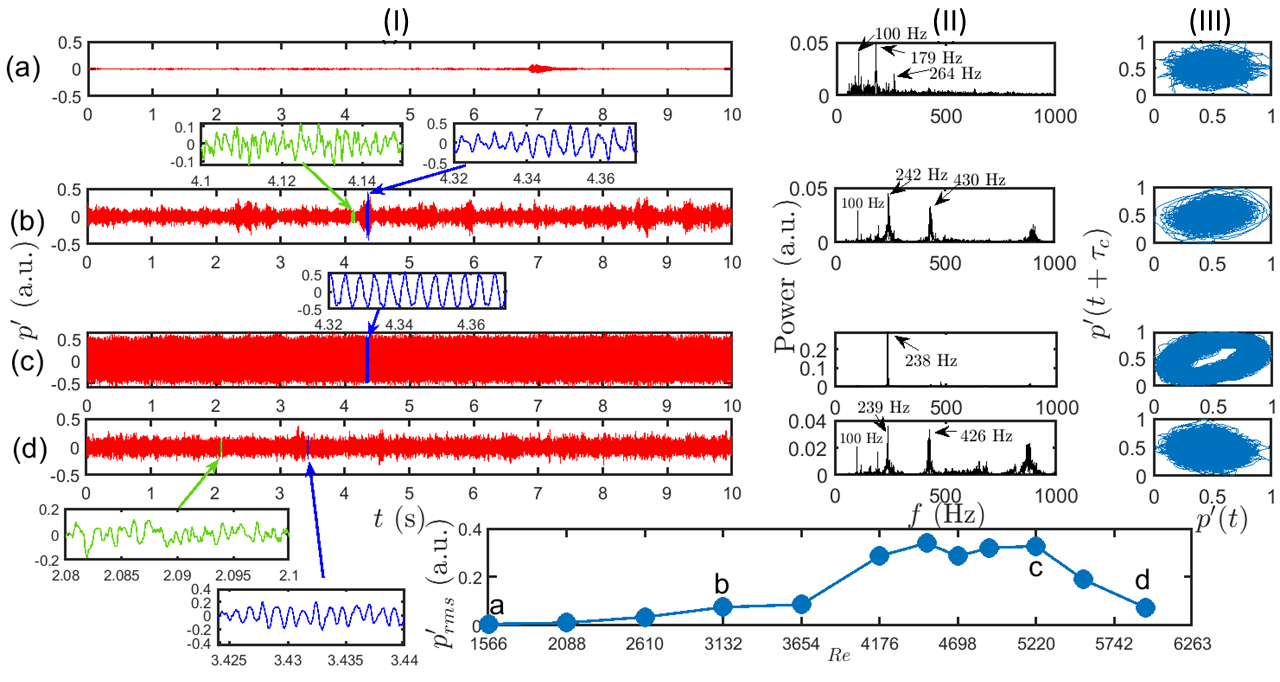}
\caption{The temporal oscillations of acoustic signal (plots (I) a-d), Fast Fourier transform of the corresponding fluctuations (plots (II) a-d) and reconstructed phase space portraits (plots (III) a-d) are presented at various flow conditions for a constant power input at 0.39 KW which are: (a) $Re$ = 1566, (b) $Re$ = 3132, (c) $Re$ = 5220, and (d) $Re$ = 5950. The root mean square of the acoustic oscillations (i.e., $p^{\prime}_{rms}$) is also shown in the figure.}\label{fig:timeseries_1}
\end{figure*}

Finally, the multifractal width ($\omega$) of spectrum $f(\alpha)$  is calculated as:
\begin{equation}
\omega = \alpha_{max} - \alpha_{min}      
\label{eq:9}
\end{equation}
 $\omega$ provides insight into the nonuniformity of the probability distribution measured on the fractal structure. Specifically, it suggests that as the value of $\omega$ increases, the multifractal features in the time series data become more complex and intense. This results in a more asymmetrical distribution and more severe fluctuations in the time series. In short, $\omega$ captures the essence of the multifractal effect within a time series, highlighting its strength and impact. 
 
 Now, let us examine the effectiveness of MFDFA measures for understanding the behavior of the IDF combustor prior to thermoacoustic instability in the following section (Sect.~\ref{sec:result}).   

\section{Result and Discussions}\label{sec:result}

In this study, we conduct experiments in two ways, as outlined in Table~\ref{Tab:1}. In Section~\ref{sec:result}, we explore the dynamics of IDF as it approaches thermoacoustic instability in both cases C-1 and C-2. To begin, we analyze the temporal evolutions and the frequency spectra of those evolutions at different dynamical states of IDF observed during the changes in control parameters (see the variation in control parameters in Table~\ref{Tab:1}). The frequency spectra and phase-space reconstructions (a detailed description of the method is provided in Supplementary Material) serve as foundational approaches to understanding the system's behavior. Additionally, we employ the 0-1 test to confirm the presence of chaotic and ordered states as combustion approaches thermoacoustic instability in the IDF configuration. We then delve into the nonlinearity associated with the system and examine the occurrence of self-similarity patterns using MFDFA. Furthermore, we test the applicability of MFDFA tools to predict the onset of thermoacoustic instability. We will discuss these aspects in detail in the following subsections. 

\subsection{Dynamics of IDF for approaching thermoacoustic instability by changing $Re$ (C-1)}

During the laminar phase ($Re \sim 1566$), we observe that the acoustic signal  ($p^{\prime}$) exhibits low amplitude aperiodic oscillations with a burst of small amplitude for a small time duration ($t \sim$ 7 seconds in Fig.~\ref{fig:timeseries_1} I a). Fast Fourier transformation of 
the temporal oscillations of the acoustic signal shows multiple significant low-frequency peaks at 100 Hz, 179 Hz, and 264 Hz(Fig.~\ref{fig:timeseries_1} II a). Sen et al.~\cite{sen2015thermoacoustic} found that the first longitudinal mode of the duct acoustics in the current IDF setup is 100 Hz, determined through an analytical solution following the approach of Dowling~\cite{dowling1995calculation} and Stow and Dowling~\cite{stow2001thermoacoustic}. Based on the methodology shown in Sen et al.~\cite{sen2018dynamic} for determining the non-isothermal and isothermal eigenfrequency modes, we can consider 100 Hz as the isothermal eigenfrequency, while 179 Hz and 264 Hz can be considered as the first and second non-isothermal eigenfrequency modes respectively.  
On the other hand, the root mean square of acoustic signal (denoted as $p^{\prime}_{rms}$) is observed to be low (as shown at point a of $p^{\prime}_{rms}$ plot in Fig.~\ref{fig:timeseries_1}) due to weak flame-acoustic coupling~\cite{sujith2020complex}. Besides, the phase-space portrait based on the acoustic fluctuations at $Re \sim 1566$ displays a strange attractor, characterized by complex combinations of phase-space subsets. This suggests that the acoustic fluctuations at such a low $Re$ can exhibit chaotic behavior.  Our previous study, using translation error~\cite{bhattacharya2021recurrence}, also found that the IDF system demonstrates high-dimensional chaos at low $Re$. We notice a similar dynamics of the system at $Re = 2088$.       

For shifting to the transitional phase from laminar ($Re \sim 3132$) while the equivalence ratio is kept fixed, we find the combination of low amplitude aperiodic oscillations and the occasional bursts of high amplitude periodic oscillations (see Fig.~\ref{fig:timeseries_1} b I). An earlier study~\cite{kasthuri2019bursting} noted similar fluctuations in the acoustic pressure of the multiple flame matrix combustor. Building on this work~\cite{kasthuri2019bursting}, we classify the fluctuations of the acoustic signal as \textquotedblleft mixed mode oscillations\textquotedblright     where we can observe one isothermal eigenfrequency mode at 100 Hz and two non-isothermal eigenfrequency modes at 242 Hz and 430 Hz (see Fig.~\ref{fig:timeseries_1} b II). On the other hand, the phase portrait of the acoustic fluctuations exhibits a distinctive structure at the periphery, indicating a significant departure in dynamical features from a chaotic system (see Fig.~\ref{fig:timeseries_1} b III). At $Re \sim 3132$, we notice that $p^{\prime}_{rms}$ increases due to the presence of occasional high amplitude bursts. We continue to notice the mixed mode behavior of the IDF combustor up to $Re = 3654$.

\begin{figure}[ht!]
\includegraphics[width=0.9\columnwidth]{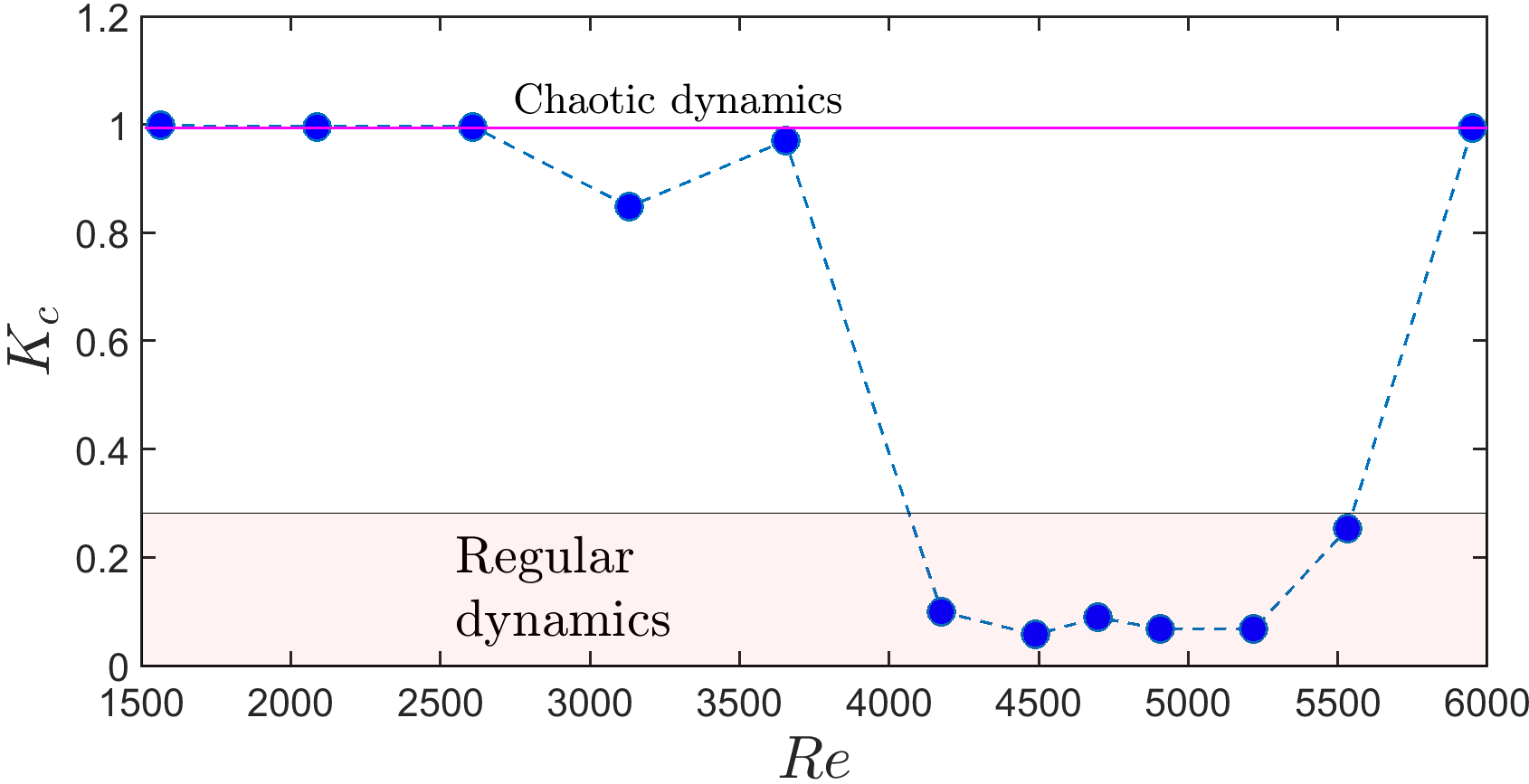}
\caption{ The performance of 0-1 test on C-1 dataset is shown in the figure. The asymptotic growth rate ($K_c$) is found to be almost 1 at $Re$ = 1566 and 2088, confirming the presence of chaotic dynamics. On the other side, $K_c$ shows lower values during $Re$ = 4176 to 5533, supporting that the system has regular dynamics. However, since the value of $K_c$ during mixed mode oscillations also shows 1 or near 1, the measure may not be helpful for understanding the subtle difference between chaotic and mixed mode state. }\label{fig:0-1test_C1}
\end{figure}

During the shifting from the transitional phase to turbulence ($Re = 3654$ to $Re = 4176 $), we observe the appearance of high amplitude periodic oscillations in the acoustic signal (see Fig.~\ref{fig:timeseries_1} c I), signifying the onset of thermoacoustic instability in the IDF combustor. This characteristic continues to be evident during $Re = 4176$ to $5533$. $p^{\prime}_{rms}$ becomes high during this period, reflecting greater fluctuations in the acoustic signal and a stronger flame-acoustic coupling in the reacting zone~\cite{sujith2020complex,kasthuri2022investigation}. In Fig.~\ref{fig:timeseries_1} c (I)-(III), we show a representative case of thermoacoustic oscillations corresponding to $Re = 5220$. During the fully developed thermoacoustic instability, we find that the non-isothernal eigenfrequency mode dominates at 238 Hz, while the other non-isothermal and isothermal eigenfrequency modes are completely suppressed (see Fig.~\ref{fig:timeseries_1} c (II) ). The phase portrait obtained during thermoacoustic instability clearly exhibits a ring-like structure, signifying the recurrence of the trajectory after a certain time interval. Such a highly deterministic periodic system is observed over a broad range of $Re$ in the turbulent flow regime of the present combustor. 

\begin{figure*}[ht!]
\includegraphics[width=2.0\columnwidth]{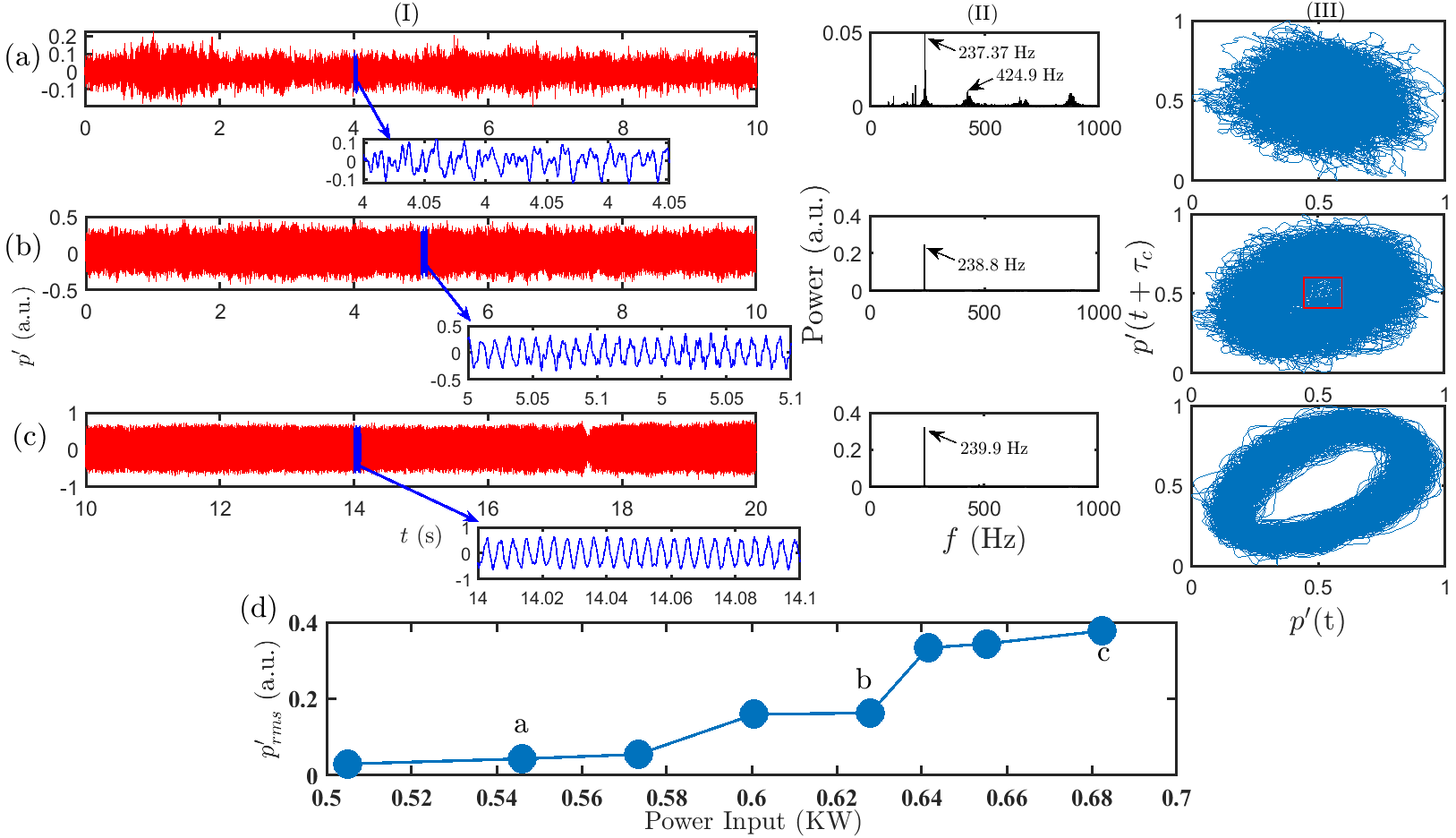}
\caption{The temporal oscillations of the acoustic signal (plots (I) a-c), Fast fourier transform of the corresponding fluctuations (plots (II) a-c) and the reconstructed phase space portraits (plots (III) a-c) are presented at following power inputs while $Re$ is almost constant: (a) $PI = 0.546$ KW, (b) $PI = 0.6279$ KW and (c) $PI = 0.6825$ KW.}\label{fig:timeseries_c2}
\end{figure*}

However, for further increasing $Re$ at 5950, we find the decline of regularity in the system. This change can be attributed to weak flame-acoustic coupling and consequently, low aperiodic oscillations appear in the acoustic signal (see Fig.~\ref{fig:timeseries_1} d I). Alongside these aperiodic oscillations, we also detect low-amplitude periodic fluctuations occurring for short time. Consequently, we see the appearance of non-isothermal eigenfrequency mode at 426 Hz in addition to the existing modes at 239 Hz and the isothermal mode at 100 Hz (see the frequency plot in Fig.~\ref{fig:timeseries_1} d I). The phase portrait, which resembles that in Fig.~\ref{fig:timeseries_1} b III, further supports the reemergence of mixed-mode dynamics at $Re = 5950$.

In short, understanding the variation in the temporal evolutions of acoustic signals obtained from the reacting flow field of IDF system and the related alteration in the frequency spectra and phase plots in accordance with the change in $Re$ (for the constant power input of 0.39 KW), we find three primary transitions within the range of $Re = 1566 - 5950 $ as follows:
\\1.Chaotic dynamics to mixed mode oscillations 
\\2.Mixed mode oscillations to limit cycle oscillations when thermoacoustic instability appears 
\\3.Limit cycle oscillations to mixed mode oscillations

To verify the difference between chaotic and regular dynamics, we conducted a 0-1 test (method is explained in Supplementary Material) on the C-1 dataset (see Fig.~\ref{fig:0-1test_C1}). An asymptotic growth rate ($K_c$) obtained from the test close to one indicates chaotic dynamics, while a value near zero suggests regular dynamics. For $Re$ ranging from 1566 to 2088, $K_c$ approaches 1, confirming the presence of chaotic dynamics. However, at $Re$ values of 2610 and 3654, where we observe mixed-mode oscillations, $K_c$ slightly deviates from 1.0. Additionally, between $Re$ values of 3653 and 5533, $K_c$ drops significantly to near zero, indicating the presence of regular dynamics or limit cycle oscillations. It is important to note that the 0-1 test is useful for distinguishing between chaotic and regular behavior in the system, but it is not very effective for understanding transitions from chaos to mixed mode dynamics.   
In the next section (Sec.~\ref{subsec:timeseries_changing PI}), we discuss the temporal evolutions of $p^{\prime}$ for the variation in power input while $Re$ is kept to be almost constant.

\subsection{Dynamics of IDF for approaching thermoacoustic instability by changing power input (C-2)} \label{subsec:timeseries_changing PI}

In Fig.~\ref{fig:timeseries_c2}, we examine the transitions in the dynamics of IDF by increasing the power input, denoted by $PI$, (i.e., by increasing the fuel flow rate) while keeping $Re$ constant  (see Table~\ref{Tab:1}). At $PI = 0.546$ KW, we observe low-amplitude aperiodic oscillations in the temporal evolution of $p^{\prime}$ (Fig.~\ref{fig:timeseries_c2} I a). From Fast Fourier transform of the acoustic signal, we find multiple non-isothermal frequency modes (see Fig.~\ref{fig:timeseries_c2} II a). The presence of multiple frequencies in the Fourier spectra, along with the strange attractor in the phase space (Fig.~\ref{fig:timeseries_c2} III a), indicates that the system exhibits chaotic behavior. Additionally, at $PI = 0.546$ KW, $p^{\prime}_{rms}$ is low, which corresponds to minimal amplitude in the acoustic sound fluctuations.

\begin{figure*}[ht!]
\includegraphics[width=2.08\columnwidth]{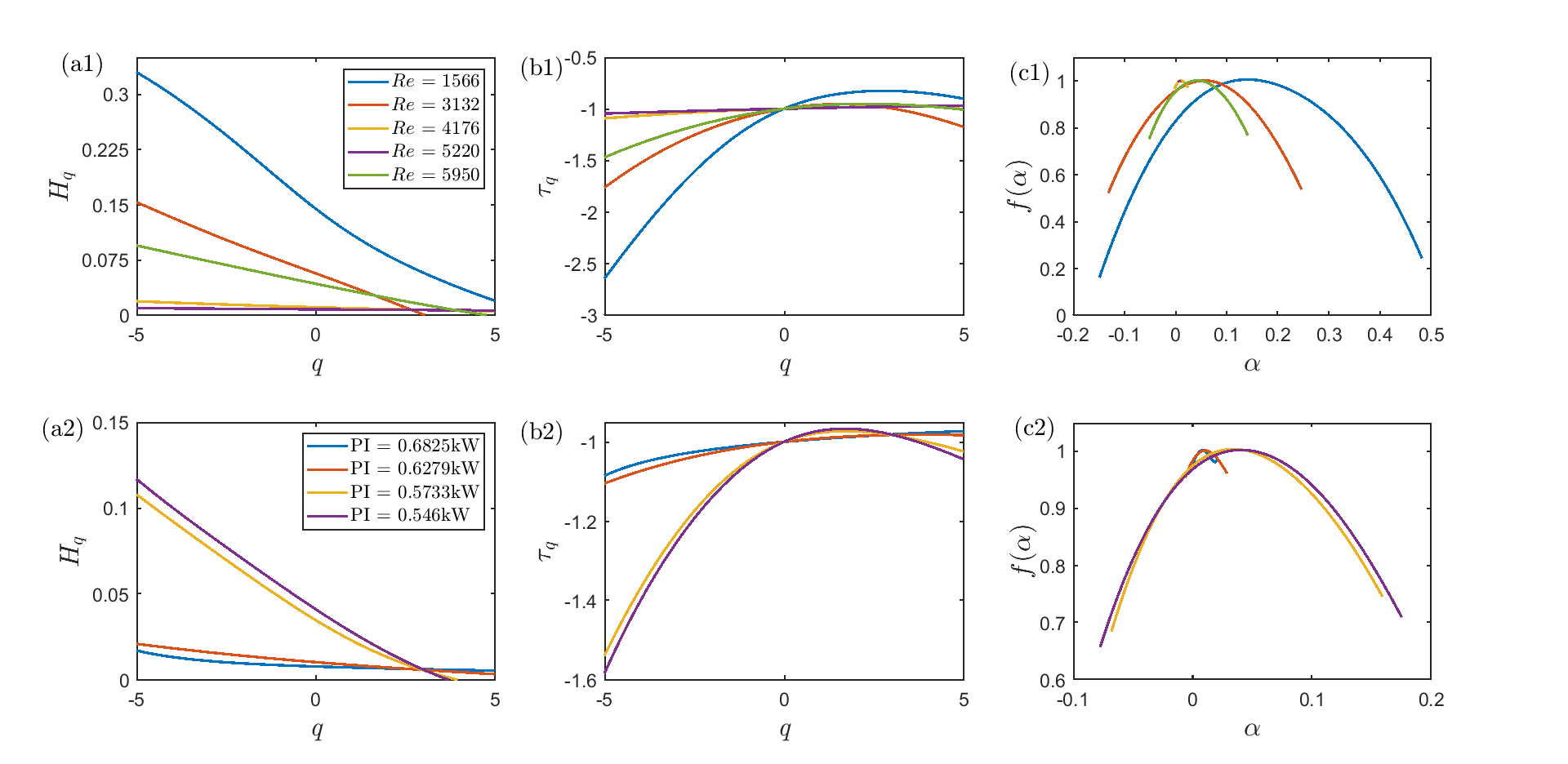}
\caption{We vary the generalized Hurst exponent ($H_q$) and multifractal scaling or mass exponent ($H_q$) as a function of $q$ for C-1 (a1 and b1) and C-2 (a2 and b2) cases. Using these two measures ($H_q$ and $H_q$), we identify the multifractal behavior, a typical characteristic of a complex state and the monofractal behavior, indication of self similar structure at different magnification of scale for different dynamical regimes of IDF subjected to the variation of $Re$ and $PI$. Further, we plot the singularity spectrum ($f(\alpha)$ over a wide range of $\alpha$ to identify the transition between multifratal and monofractal behavior of the system (see plots c1 and c2).}\label{fig:mutlifractal}
\end{figure*}

At $PI = 0.6279$ KW, the IDF system begins to display organized behavior, as the temporal evolution of $p^{\prime}$ shows an almost periodic oscillation. Additionally, the corresponding Fourier spectra reveal a prominent single peak near 239 Hz, indicating a non-isothermal eigenfrequency mode (Fig.~\ref{fig:timeseries_c2} II b). 
In the center of the phase phase (marked by red square), we observe an obscure void where the ring-like structure is less prominent under this condition (Fig.~\ref{fig:timeseries_c2} III b). However, we can discern that a limit cycle attractor emerges, while a strange attractor remains present in the phase space. On the other hand,  $p^{\prime}_{rms}$ at $PI = 0.6279$ KW becomes higher as the fluctuations of acoustic sound appears as higher amplitude than that of $PI= 0.546$ KW. Overall, we understand that flame-acoustic coupling at $PI=0.6279$ is stronger than that evidenced for chaotic state and weaker than that can be seen during limit cycle oscillations. Thus, we characterize this state as a \textquotedblleft weakly ordered\textquotedblright  state. We observe similar behavior in the combustor at $PI = 0.6$ KW. 

For further increase in $PI$ at 0.6825 KW, we find the periodic behavior having high amplitude in temporal oscillations of $p^{\prime}$ (Fig.~\ref{fig:timeseries_c2} I c). The Fourier spectra show a high-power single dominant peak at 239 Hz (non-isothermal eigenfrequency mode), and the limit cycle attractor in the phase space confirms the presence of ordered dynamics (Figs.~\ref{fig:timeseries_c2} II c and~\ref{fig:timeseries_c2} III c respectively). $p^{\prime}_{rms}$ at $PI = 0.6825$ KW becomes even higher than that of  $PI= 0.6279$ KW. An 0-1 test is also performed based on the C-2 dataset to confirm the dynamical transitions from chaotic to order (See Supplementary Material).

In a nutshell, it is crucial to understand that as the operating conditions for both the C-1 and C-2 cases lead the combustor's dynamics towards thermoacoustic instability, the non-isothermal eigenfrequency mode, which is nearly 238 Hz, becomes dominant, while other modes are significantly diminished. Therefore, a gradual increase in the power of this frequency mode can serve as an indicator of thermoacoustic instability for this combustor. In the next section (Sec.~\ref{subsec:multifractal analysis}), we characterize the dynamical changes during the transitions to thermoacoustic instability in IDF combustor using MFDFA approach.

\subsection{Multifractal analysis: Generalized Hurst exponent, mass exponent and multifratal spectrums for (C-1) and (C-2)} \label{subsec:multifractal analysis}

Following the MFDFA approach, we vary the structure-function ($F_q$)  to account for the changes in the measurement scale ($S$) and estimate the slope of $F_q$ vs. $S$ for different values of the order of fluctuations ($q$). The resulting slope for different values of $q$, known as the generalized Hurst exponent (noted by $H_q$, as mentioned in Sec.~\ref{sec:method}), is discussed in detail in the appendix. As previously mentioned, the slope for $q=2$ is popularly known as the Hurst exponent, which is discussed later. In Fig.~\ref{fig:mutlifractal},  we observe the trends of $H_q$ (see plots a1 and a2) and mass exponent $\tau_q$ (see the plots b1 and b2) that measures the degree of multifractality for different  order of fluctuations $q$. In addition, we try to see the multifractal spectrum which illustrates the variation of $f(\alpha)$ with $\alpha$ (see plots a1 and a2 in Fig.~\ref{fig:mutlifractal}). From Fig.~\ref{fig:mutlifractal} a1, we notice that $H_q$ becomes higher for -5 $\le q \le 0$ than for 0 $\le q \le 5$ during $Re = 1566 - 2088$ (chaotic regime), indicating that the small-scale oscillations in the system dominate over large-scale fluctuations, a typical characteristic of multifractality. Moreover, the nonlinear relationship of $\tau_q$ with $q$ across its range confirms a higher degree of multifractality~\cite{tanna2014multifractality, ZhanOptimal} in the acoustic signal. Consequently, we observe that the singularity spectrum $f(\alpha)$ displays a bell-shaped variation over a wide range of singularity strength $\alpha$.  

We notice a similar nonlinear dependence of $H_q$ (characterized by convex lines) for the variation of $q$ during $Re = 2610-3654$ (mixed mode dynamics), suggesting that the time series representing the mixed mode dynamics of the system is also multifractal.  However, the values of $H_q$ during -5 $\le q \le 0$ are lower than those observed during chaotic dynamics. The loss of degree in multifractality during mixed mode dynamics is reflected in the variation of $\tau_q$ with $q$ as the nonlinear dependence is found only within a specific range of $q$ (see $Re = 3132$ in Fig.~\ref{fig:mutlifractal} b1). Consequently, the span of $\alpha$ over which $f(\alpha)$ varies reduces (see $Re = 3132$ in Fig.~\ref{fig:mutlifractal} c1).

On the other hand, as the system exhibits regular dynamics during thermoacoustic instability ($Re = 4176-5533)$, the trend of $H_q$ becomes almost linear and independent of $q$ (see $Re =$ 4176 and 5220 in Fig.~\ref{fig:mutlifractal} a1), indicating that the temporal variation of system becomes monofractal in nature. The monofractal behavior emerges when a system exhibits a similar pattern although the scale of the measurement is magnified, signifying the presence of self-similarity in behavior. Consequently, $\tau_q$ becomes linear over the range of $q$, and the collapse of $f(\alpha)$  confirms the diminish of the multifractality in the system. Again, as the system returns back to the mixed mode regime as a consequence of the increment in $Re$ to 5950, we notice that $H_q$ is not constant over the range of $q$, signifying the reappearance of multifractal behavior of the time series. In a nutshell, we notice the following transitions in the dynamics of the system for increase in $Re$:
multifractal (chaotic region) $\rightarrow$ loss of multifractality (mixed mode dynamics) $\rightarrow$ monofractal (fully developed thermoacoustic instability) $\rightarrow$ multifractal (mixed mode dynamics). 

In Figs.~\ref{fig:mutlifractal} a2,b2 and c2, we observe the variation of $H_q$, $\tau_q$ for -5 $\le q \le 5$ and the variation of $f(\alpha)$ with $\alpha$ for C-2 dataset to distinguish the multifractal and monofractal nature of the system with respect to the variation of $PI$. We notice that the temporal variations of the system exhibit a multifractal nature when $PI$ lies between 0.505 KW and 0.5733 KW since $H_q$ and $\tau_q$ are not constant over the range of $q$. Consequently, the variation of the singularity spectrum function exhibits a bell-shaped curve over a wide span of $\alpha$. However, as the system enters into the weakly ordered state (during $PI$ = 0.6 to 0.6279 KW), $H_q$ and $\tau_q$ become almost constant for -5 $\le q \le 5$ due to the emergence of regular dynamics. As a result, the variation of $f(\alpha)$ with $\alpha$ shows a significant shrink of the spectrum width. Further, the width of the spectrum progressively reduces when the combustor becomes susceptible to thermoacoustic instability (see $PI$ = 0.6826 KW in Figs.~\ref{fig:mutlifractal} c2). Thus, for the C-2 case, during the range of $PI$ = 0.505 KW to 0.6825 KW, we observe the following transition of the dynamics:
multifractal (chaotic region) $\rightarrow$ loss of multifractal (for weakly ordered) $\rightarrow$ monofractal (during fully developed thermoacoustic instability). 

\begin{figure*}[ht!]
\includegraphics[width=2.08\columnwidth]{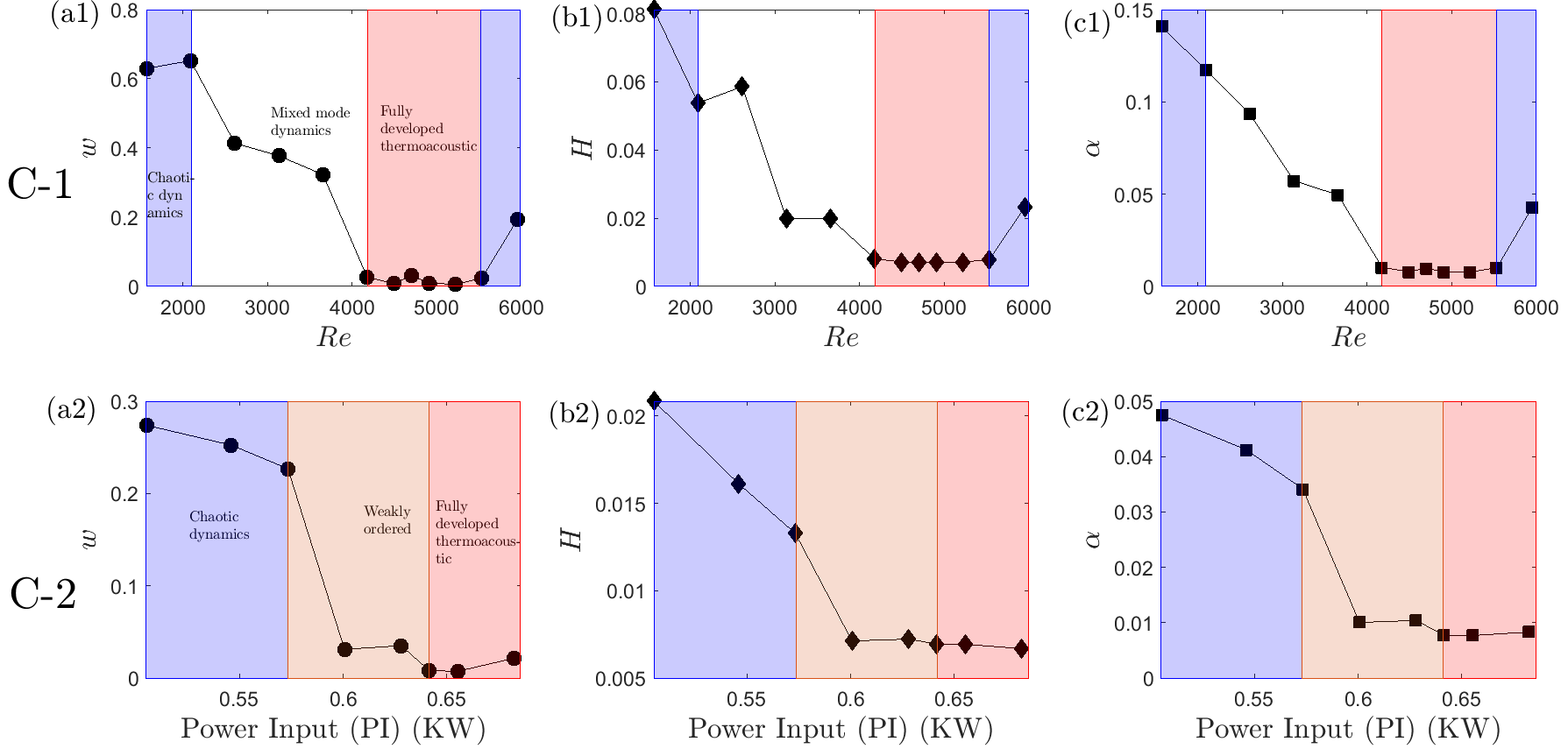}
\caption{We estimate the multifractal width (denoted here by $\omega$), Hurst exponent ($H$) and the singular strength (denoted by $\alpha$) for both C-1 and C-2 processes. The variation of $\omega$ (a1 and a2), $H$ (b1 and b2) and $\alpha$ (c1 and c2) are shown here.  }\label{fig:mutlifractalmeasures}
\end{figure*}

\subsection{Measures to distinguish different dynamical states: multifractal width, Hurst exponent, singularity exponent} \label{subsec:multifractal measures}

In Fig.~\ref{fig:mutlifractalmeasures}, we estimate two MFDFA measures, namely, multifractal spectrum width (denoted by $\omega$ in the plots a1 and a2) and singular strength ($\alpha$ in the plots c1 and c2) and one DFA (detrended fluctuation analysis, where single factual dimension is used) measure such as Hurst exponent ($H$ in the plots b1 and b2) for the variation in $Re$ and $PI$ in C-1 and C-2 cases, respectively. The purpose of these calculations is to examine how these measures vary across different dynamical regimes observed in IDF. Additionally, the investigation aims to assess the effectiveness of these tools in distinguishing between the dynamical regimes within this specific configuration of fuel-air combustion.

Let us first look at the variations of $\omega$ for the alteration in $Re$ (Fig.~\ref{fig:mutlifractalmeasures} a1) and $PI$ (Fig.~\ref{fig:mutlifractalmeasures} a2). We find that $\omega$ becomes higher for the chaotic dynamics seen during $Re = 1566-2088$ and $PI =0.505-0.5733$. During mixed mode oscillations ($Re = 2610-3654$ in case C-1), $\omega$ drops from around 0.6 to 0.3. On the other hand, the value of $\omega$ in case C2 during the chaotic region is seen between 0.21 and 0.29. However, the trend of $\omega$ in both cases drops as the flame dynamics deviates from the chaotic state. Interestingly, the measure significantly drops well below 0.1 during the weak limit cycle oscillations region for the C-2 case. During fully developed thermoacoustic instability or limit cycle oscillations, we notice near-zero values of $\omega$ for both C-1 and C-2 cases. Thus, $\omega$ can be a promising measure for identifying the transitions between the states as the continuous decreasing trend of this measure is seen till the onset of thermoacoustic instability. Also, we notice an increase in the measure when the system returns back to the mixed mode regime in the C-1 case.
It is important to note that the thresholds for identifying transitions between chaotic and mixed mode and between chaotic and weak limit cycle regions differ in the two processes. However, there is a significant drop in the measure during these transitions—from chaotic to mixed mode in the C-1 case and from chaotic to a weakly ordered state in the C-2 case. This notable change in the measure suggests a promising trend for understanding inter-state switching. Additionally, a substantial decrease to nearly zero during the thermoacoustic instability in both processes indicates that $\omega$ is a suitable predictor for the onset of this instability, regardless of the previous state (mixed mode or weak coupling).

\begin{figure*}[ht!]
\includegraphics[width=1.3\columnwidth]{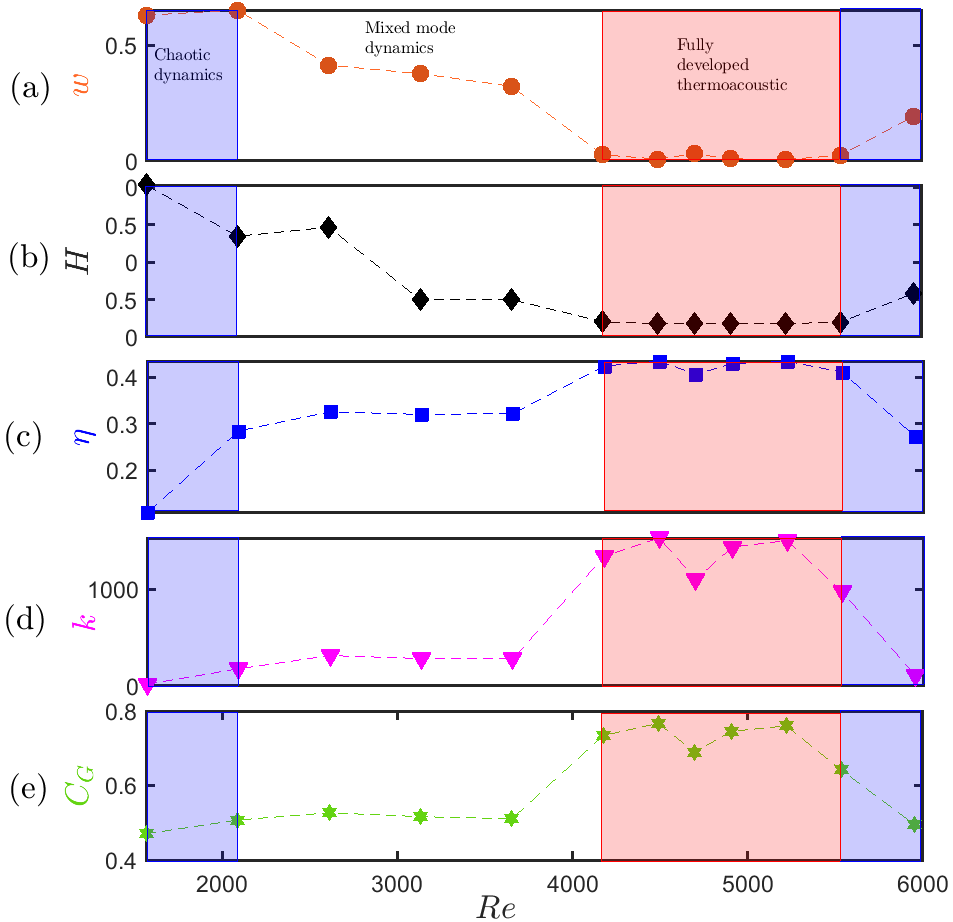}
\caption{We perform a recurrence network analysis, as shown in Bhattacharya et al.~\cite{bhattacharya2021recurrence}. Here, we reproduce the variations of the network parameters, namely, global efficiency ($\eta$), degree centrality ($k$) and global clustering coefficient ($C_G$) by estimating them on C-1 (source:Chaos 31, 033117 (2021), with a permission from AIP Publishing)~\cite{bhattacharya2021recurrence}. Further, we compare these network measures with $H$ and $\omega$. We show the comparison using dataset C-1 only to avoid the repetition.}\label{fig:compare_measures_C1}
\end{figure*}

Next, we check the variation of $H$ for different dynamical states in Figs.~\ref{fig:mutlifractalmeasures} b1 (for the variation in $Re$) and ~\ref{fig:mutlifractalmeasures} b2 (for the variation in $PI$). We find $H \le 0.1$ over all the dynamical states in both cases (C-1 and C-2), indicating the long-term switching~\cite{alvarez2008time} of the temporal evolutions of the acoustic fluctuations of the system between the higher and lower values. Thus, the trend of the temporal evolutions of the acoustic fluctuations becomes anti-persistent~\cite{ihlen2012introduction} throughout the operating regimes. However, in the current study, our prime interest is to observe the trend of the variation of $H$ at the transitions between the dynamical states. We find a sharp decrease in the measure during the transition from chaotic to mixed mode (C-1 case) and chaotic to weakly ordered state (C-2 case). Further, we find a slow decline in the measure while the system undergoes a transition from mixed mode to thermoacoustic state (C-1 case) and weakly ordered to thermoacoustic state (C-2 case). Similar to $\omega$,  the variation of $H$ during the switching between inter-states is found promising. In addition, the gradual decline in $H$ can be suitable to predict the onset of thermoacoustic instability in IDF combustor, as similarly found for the other turbulent combustors~\cite{pavithran2020universality,nair2014multifractality}. However, note that the change in $H$ from mixed mode to thermoacoustic is noticed to be more prominent than that from weakly ordered to thermoacoustic state. An almost similar trend of $\alpha$ is noticed in Figs.~\ref{fig:mutlifractalmeasures} c1 and c2.      

In a nutshell, the decreasing trends of $\omega$, $H$ and $\alpha$ prior to the thermoacoustic instability can be useful for identifying the boundaries between the states. However, the threshold for defining these boundaries relies on the control parameters, as the measured values at transitions are not consistent for the two processes. In the next section (Sec.~\ref{compare with existing measures}), we perform a comparative analysis of the proposed measures with existing methods used for IDF in the previous literature~\cite{bhattacharya2021recurrence}.

\subsection{Robustness of the properties of multifractal properties for thermoacoustic instability: Comparison with the existing measures}
\label{compare with existing measures}

In this section, we use a complex network approach based on the concept of recurrence, which is popularly known as a recurrence network. We estimate the statistical measures of the network, which can capture the intrinsic properties of the systems at different states. Further, we compare these network measures with $H$ and $\omega$ across various dynamical states of IDF.  

The construction of the recurrence network involves two main steps: first, a phase space reconstruction, and second, the creation of a network based on an adjacency matrix. This adjacency matrix is derived from the recurrence matrix after subtracting the Kronecker delta to eliminate self-loop connections. A more detailed discussion on the network construction is available in the Supplementary Material.  We estimate three network parameters, as  outlined by our previous investigation~\cite{bhattacharya2021recurrence}, which are: global efficiency ($\eta$), degree centrality ($k$) and global clustering coefficient ($C_G$). The mathematical formulations for these three measures are provided in the Supplementary Material.

\begin{figure*}[ht!]
\includegraphics[width=2\columnwidth]{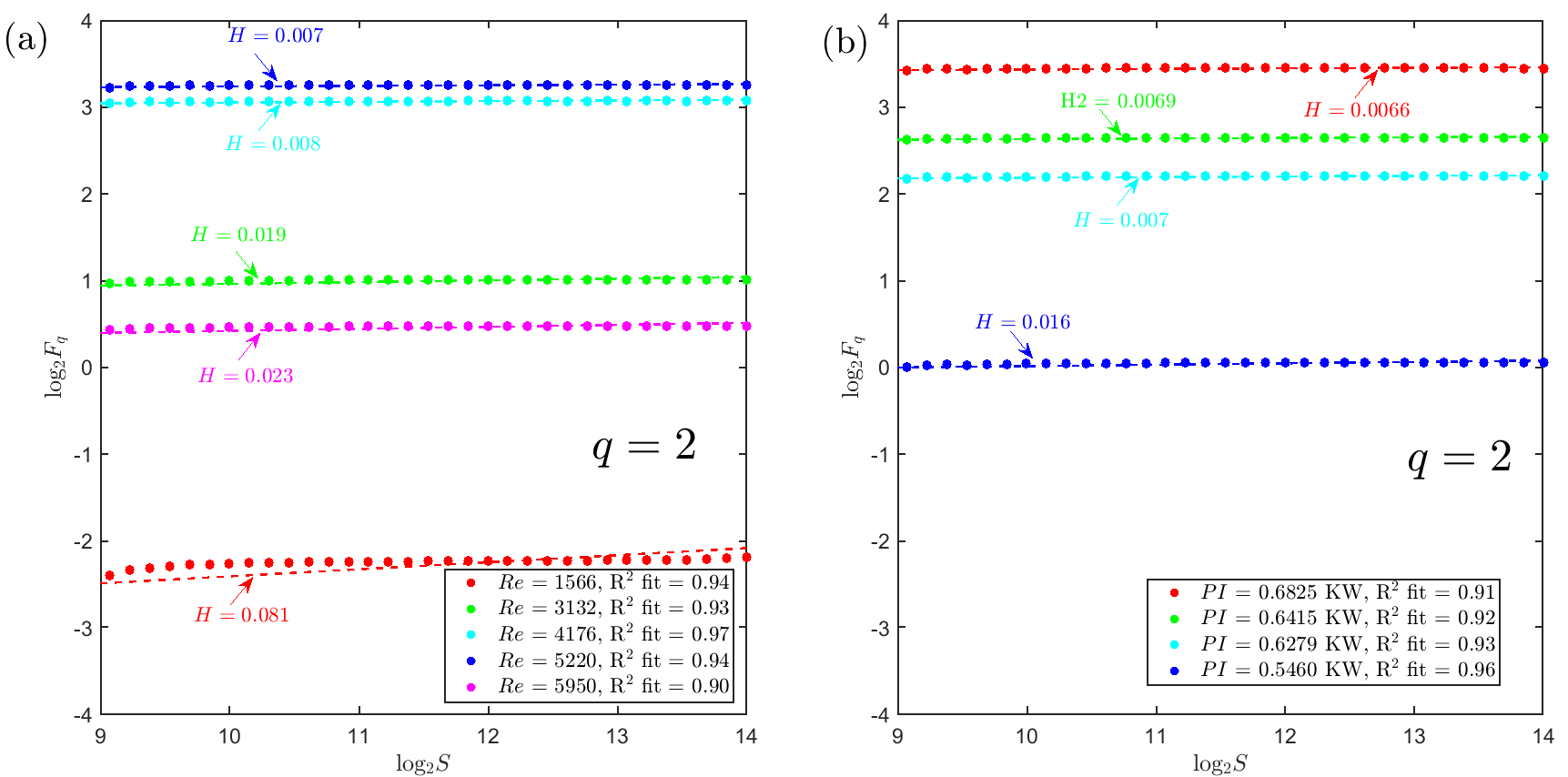}
\caption{The variation of structure-function ($F_q$) is shown here for the range of scale ($S$) for C-1 and C-2 cases in (a) and (b) respectively. In (a) and (b), we show the Hurst exponent ($H$) from the slope obtained from log-log plot of $F$ vs $s$ as the order of fluctuations ($q$) is considered as 2 here.   }\label{fig:Structure_function_q_2}
\end{figure*}

In Fig.~\ref{fig:compare_measures_C1}, we show the performance of the network measures and conduct a comparative study between these well-known network measures and MFDFA-DFA measures on C-1 dataset. We notice a dinstinct variation in $\omega$ during the transition from chaotic to mixed mode oscillations (from $Re =$ 2088 to 2610), while the variation is not prominent in $H$ and the network measures ($k$, $\eta$, $C_G$). However, both $H$ and $\omega$ show significant variation corresponding to sudden bursts in the acoustic oscillations during mixed-mode conditions. On the other hand, no notable variation is seen in the network measures during the mixed-mode oscillations. Nonetheless, both network and MFDFA measures exhibit a prominent switch when the flame dynamics changes from the mixed mode to the thermoacoustic instability state. Notably, the MFDFA-DFA measures display a stable trend once fully developed thermoacoustic conditions are established in the combustor, while the network measures show some fluctuations. Overall, MFDFA-DFA measures prove to be better indicators for identifying inter-state switching, although the trends of the network measures are promising, particularly at the onset of instability.  

Thus, the current investigation utilizing the MFDFA approach provides valuable insights into the transitions of the IDF combustor that lead to thermoacoustic instability. The characterization of various combustion states elucidated in the study can further inspire researchers in physics-inspired learning on the critical transitions of both laminar and turbulent flames.    

\section{Conclusions}\label{sec:conclusions}

In this study, we aim to characterize the transitions of an inverse diffusion flame (IDF) that lead the system toward thermoacoustic instability using a well-established method known as multifractal detrended fluctuation analysis (MFDFA). We conduct our investigation by manipulating the control parameters in two ways: 
1. Varying the flow Reynolds number while keeping the global equivalence ratio constant (referred to as case C-1).
2. Changing the power input (i.e., flue flow rate) while maintaining a nearly constant Reynolds number (referred to as case C-2).
Through these approaches, we examine all potential transitions that the IDF experiences at a fixed flame location. In our findings, we identify transitions in the IDF as the flow conditions shift from a laminar phase to a turbulent phase in case C-1. We observe a pathway from a chaotic state to an ordered state, passing through a mixed-mode oscillation state in this case. Conversely, in case C-2, we notice the transitions of the IDF shifting from a chaotic state to an ordered state via a weakly coupled state. 

To gain deeper insights into these transitions, we employ MFDFA tools. Our analysis reveals a higher degree of multifractality, evidenced by a broad multifractal spectrum, during the chaotic state for both the C-1 and C-2 cases. However, a loss in multifractality is observed as the flame transitions from a chaotic state to mixed mode oscillations in the C-1 case, or to a weakly coupled state in the C-2 case. The nature of multifractality diminishes completely, with the multifractal spectrum narrowing and a monofractal nature becoming dominant when thermoacoustic instability emerges in the IDF.
The MFDFA performance in this study also evaluates the effectiveness of the Hurst exponent and MFDFA tools, such as spectrum width and singularity strength. We find that these three parameters are effective in defining the dynamical transitions of the IDF. Therefore, we recommend these measures not only as early indicators of thermoacoustic instability but also as reliable indicators of dynamical transitions.

The current study performs a holistic investigation on the possible transitions of IDF. However, in future, we plan to develop a model that accurately represents the physics-inspired machine learning approaches to predict the dynamical transitions obtained at different flame locations at real-time.  
\begin{figure*}[ht!]
\includegraphics[width=2\columnwidth]{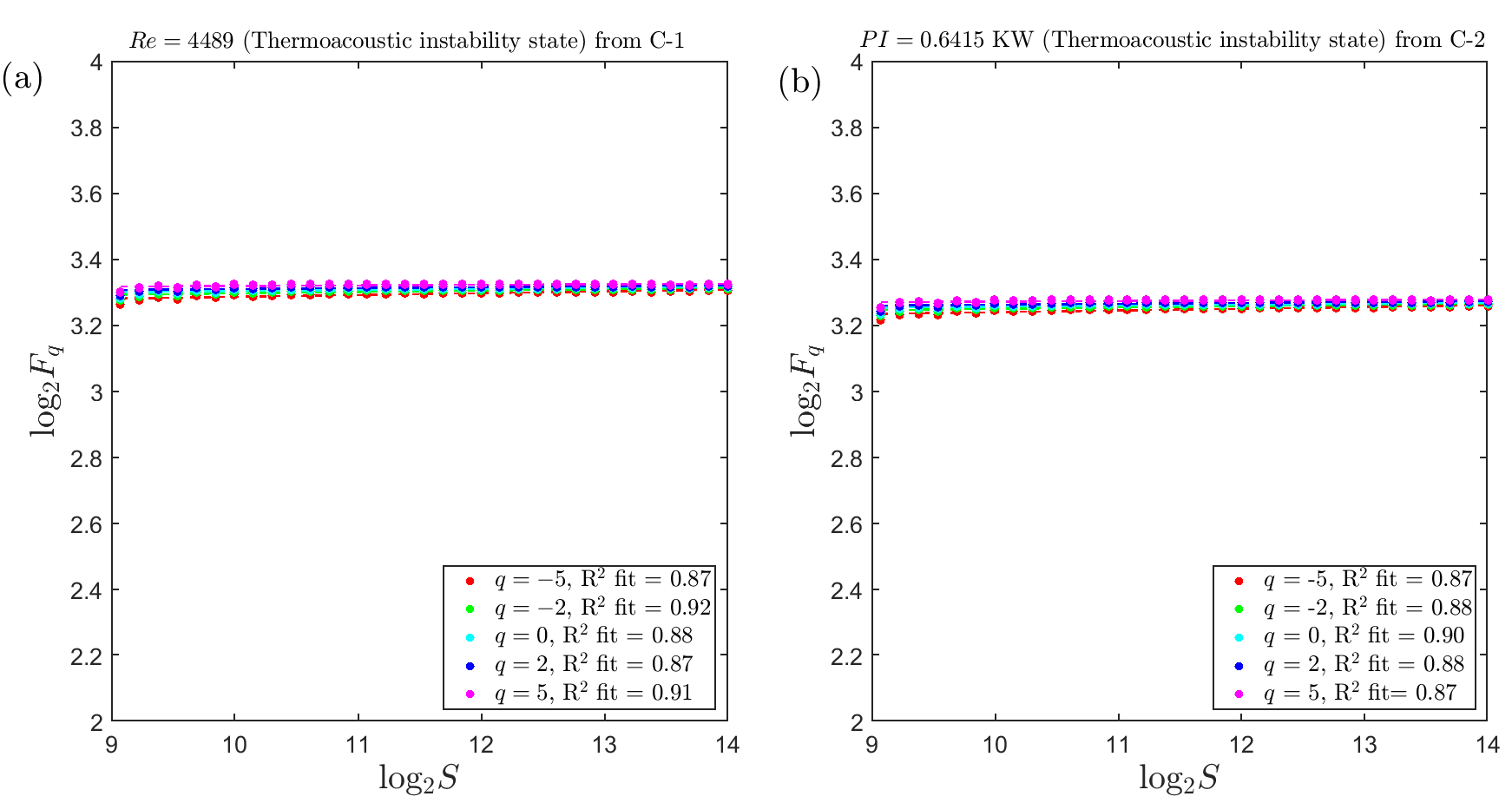}
\caption{The variation of structure-function ($F_q$) is shown here for the range of scale ($S$) during thermoacoustic instability state ($Re = 4489$ from C-1 and $PI= 0.6415$ KW from C-2 case shown as the representative) in (a) and (b). In (a) and (b), we show the Generalized Hurst exponent ($H_q$), which is the slope obtained from a log-log plot of $F$ vs $s$ for different orders of fluctuations ($q$).}\label{fig:Structure_function_diff_q}
\end{figure*}

\begin{figure*}[ht!]
\includegraphics[width=2\columnwidth]{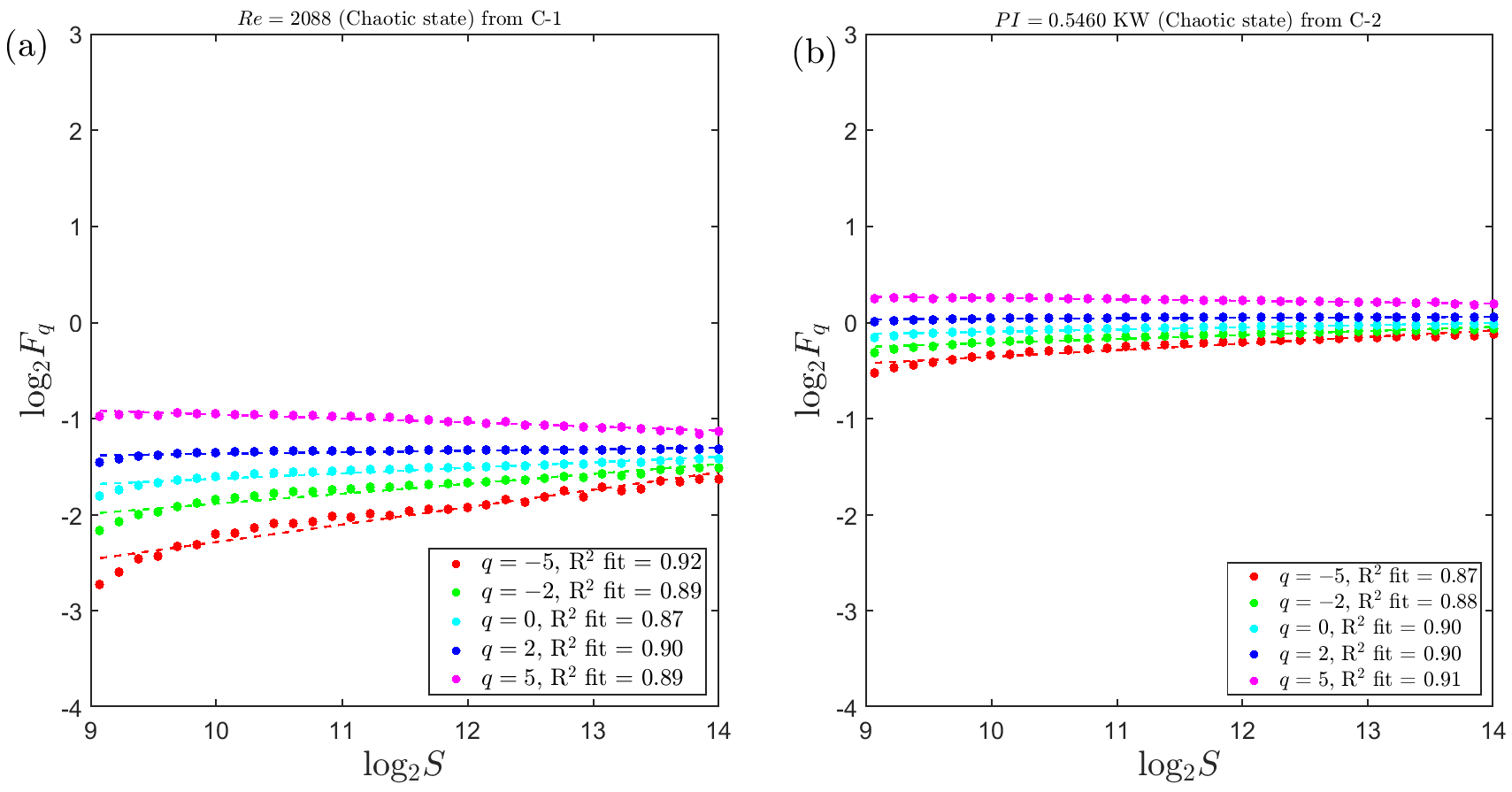}
\caption{The variation of structure-function ($F_q$) is shown here for the range of scale ($S$) during chaotic state ($Re = 2088$ from C-1 and $PI= 0.5460$ KW from C-2 case shown as the representative) in (a) and (b). In (a) and (b), we show that $H_q$, which is the slope obtained from a log-log plot of $F$ vs $s$, varies for the change in order of fluctuations ($q$).}\label{fig:Structure_function_diff_q_chaos}
\end{figure*}

\section*{Supplementary Material}

We include Supplementary Material that clearly outlines the methods of phase reconstruction, recurrence networks, and the 0-1 test. The performance of the 0-1 test for the C-2 case is also detailed in the Supplementary Material.

\section*{Acknowledgements}
S.D. wants to thank Prof. R I Sujith for his guidance on combustion instabilities. S.D. also wants to thank the combustion lab at Jadavpur University for performing experiments and using lab accessories.

\section*{Data Availability Statement}

The data that support the findings of this study are available from the corresponding author upon reasonable request.

\section*{Declarations: Conflict of interests}
The authors declare no competing interests.

\section*{Appendix}\label{sec:appendix}
\subsection{Statistics of multifractal analysis}

\subsubsection{Structural function vs scale for different states for q=2 for (C-1) and (C-1)}

In Figs.~\ref{fig:Structure_function_q_2} a and b, we present the logarithm plots of structure-function (denoted by $F_q$) with the variation in the scale of measurement (denoted by $S$) when the order of fluctuations ($q$) is considered as 2. We observe that $H$, slop of log-log plot of $F_q$ vs. $S$ for $q=2$, becomes lower compared to that obtained for the chaotic state (low $Re$ in C-1 or low $PI$ in C-2 case) once IDF becomes susceptible to the thermoacoustic instability.  

\subsubsection{Structural function vs scale for thermoacoustic state for different q for (C-1) and (C-2)}

In Figs.~\ref{fig:Structure_function_diff_q} (a) and (b), we present $F_q$ vs $S$ for different orders of fluctuations ($q$) during thermoacoustic instability. We present the variation of $F_q$ with $S$ at $Re = 4489$ in Fig.~\ref{fig:Structure_function_diff_q}(a)  and at $PI =0.6415$ KW in Fig.~\ref{fig:Structure_function_diff_q}(b). We confirm that the slope of the plots during thermoacoustic instability remains exactly the same, irrespective of the change in $q$.

\subsubsection{Structural function vs scale for Chaotic state for different q  for (C-1) and (C-2)} 


In Figs.~\ref{fig:Structure_function_diff_q_chaos} (a) and(b), we present $F_q$ vs $S$ for $q$ during chaotic state. We present $F_q$ vs $S$ at $Re = 2088$ (from C-1 case) in Fig.~\ref{fig:Structure_function_diff_q_chaos}(a) and at $PI =0.546$ KW (from C-2 case) in Fig.~\ref{fig:Structure_function_diff_q_chaos}(b). We notice that the slope of the plots during the chaotic state changes for the change in $q$.

\bibliography{manuscript_1}

\end{document}